\documentclass{article}
\usepackage[preprint]{neurips_2026}

\usepackage[utf8]{inputenc}
\usepackage[T1]{fontenc}
\usepackage{hyperref}
\usepackage{url}
\usepackage{booktabs}
\usepackage{amsfonts}
\usepackage{amsmath}
\usepackage{nicefrac}
\usepackage{microtype}
\usepackage{graphicx}
\usepackage{xcolor}
\usepackage{multirow}
\usepackage{subcaption}
\usepackage{tablefootnote}
\usepackage[most]{tcolorbox}
\usepackage{enumitem}
\usepackage[table]{xcolor}
\usepackage{svg}
\graphicspath{{figures/}}
\usepackage{fancyvrb}

\definecolor{rowgray}{gray}{0.94}
\definecolor{overallgray}{gray}{0.87}
\tcbuselibrary{listings,breakable}

\title{Anatomy of a Query: \\
W5H Dimensions and FAR Patterns for Text-to-SQL Evaluation}

\author{Vicki Stover Hertzberg, Eduardo Valverde, Joyce C. Ho \vspace{3mm}\\
Emory University\\ 
Georgia Institute of Technology, Atlanta, USA \\
\texttt{vhertzb@emory.edu} \quad
\texttt{evalverde3@gatech.edu} \quad
\texttt{joyce.c.ho@emory.edu}
}

\begin{document}

\maketitle

\begin{abstract}

Natural language interfaces to databases have gained popularity, yet the theoretical foundations for evaluating and designing these systems remain underdeveloped. We present QUEST (Query Understanding Evaluation through Semantic Translation), a framework resting on two independently motivated components: the FAR structural invariant, which holds that every well-formed query reduces to Filter, Aggregate, and Return operations; and the W5H dimensional framework, which holds that all filtering criteria map to six semantic dimensions (Who, What, Where, When, Why, and How). Validated across five text-to-SQL datasets (n = 120,464), FAR conformance is universal across all domains and schema types, while W5H dimensional profiles vary substantially. Healthcare queries are strongly concentrated in temporal (WHEN: 80.4\%) and person-centric (WHO: 73.0\%) dimensions far exceeding general-domain benchmarks, and causal (WHY) and mechanistic (HOW) reasoning are near-zero everywhere, with apparent HOW exceptions reflecting quantitative aggregation rather than genuine procedural reasoning. These results identify a frontier that must be crossed for genuine machine reasoning over structured data.

\end{abstract}

\section{Introduction}

Early in the modern era of computation, programmers wrote code caring only that it was accurate and efficient, making incremental edits until acceptable tolerance limits were reached.
In the late 1960s, two developments brought systematic discipline to programming.
First, B\"ohm and Jacopini's structured program theorem~\citep{BohmJacopini66} demonstrated that all computable functions result from just three control structures: sequence, selection, and iteration.
Second, Edsger Dijkstra's influential call for `structured programming'~\citep{dijkstra68} transformed ad-hoc coding practices into principled software development.

Today, database query understanding faces a similar challenge.
Applications that use natural language to query databases have 
exploded since the 2022 breakthroughs in large language models \citep{ouyang2022training}. 
Yet evaluation of these systems has remained 
anchored to two metrics: execution accuracy, which tests whether a predicted query 
returns the same result set as the gold standard, and exact set match, which 
tests structural identity. Both metrics measure average performance on known 
distributions without characterizing the structural or semantic operations 
a system must handle correctly. The consequences are severe: models achieving 86\% 
execution accuracy on standard benchmarks drop to 6\% on enterprise workflows 
\citep{lei2025spider20evaluatinglanguage}, and systematic re-evaluation finds that exact matching 
fails consistently due to query underspecification and reference 
query ambiguity \citep{pourreza2023evaluating}. A system can score well on 
general-domain benchmarks and fail catastrophically in deployment, and current 
metrics provide no principled way to diagnose why or predict where the next failure will occur.

In this paper, we present \textbf{QUEST} (Query Understanding Evaluation 
through Semantic Translation), a principled framework for database 
information retrieval built on two co-equal pillars. The first is the 
\textbf{W5H dimensional framework}, which holds that all criteria for a 
database query comprise one or more of the following questions: Who? What? 
When? Where? Why? How? The second is the \textbf{FAR structural invariant}, 
which holds that all database queries reduce to three fundamental steps: 
\textbf{filter} on specific criteria, \textbf{aggregate} all records that 
meet those criteria, and \textbf{return} those records for further analysis. 
These two pillars are unified by the Filtered Enumeration Principle: every 
query reduces to a command to list, count, or aggregate the records meeting 
a W5H dimensional specification. Together, W5H and FAR provide a universal 
grammar for query construction and evaluation.

Section~\ref{sec:theory} describes the theoretical foundation of QUEST, 
illustrating both pillars with case studies of increasing complexity. 
Section~\ref{sec:empirical} presents an empirical validation across five 
text-to-SQL datasets spanning 120,464 queries. Section~\ref{sec:implication} 
describes the framework's implications for system design and evaluation. 
Section~\ref{sec:related} situates QUEST within related work. Section~\ref{sec:conclusion} summarizes the contributions and identifies directions for future work.
\section{Theoretical Foundation for QUEST}
\label{sec:theory}

Text-to-SQL systems fail not because queries are unpredictable, but because the principles underlying them have never been made explicit. QUEST addresses this by applying a reductive insight analogous to structured programming. Just as all computable functions reduce to a small set of control structures, all database queries reduce to two co-equal pillars: FAR and W5H, unified by the Filtered Enumeration Principle.

\subsection{The W5H Dimensional Framework}
\label{sec:w5h_theory}

Human information needs have a natural structure. Although originally developed in the context of communication and cognitive categorization respectively, Lasswell's communication model \citep{lasswell1948structure}  and Rosch's categorization theory \citep{Rosch1978} both converge on the same six dimensions, WHO, WHAT, WHERE, WHEN, WHY, and HOW as the natural axes along which people organize what they want to know. QUEST draws on this convergence as its semantic foundation. Rather than imposing an external decomposition on database queries, it surfaces the structure users are already thinking in, thereby reducing the translation burden between user intent and system input.

Every natural language query that can be answered by a database implicitly invokes one or more of these six dimensions. The W5H framework makes that implicit structure explicit, providing the semantic middle layer through which text maps to SQL. A user asking `which flights from Boston depart before noon?' is specifying WHERE and WHEN whether they know it or not; the SQL that answers it must encode exactly those dimensions. Formally, we define each dimension as follows:

\begin{itemize}

\item \textbf{WHO} refers to the primary entities represented by the records (e.g., people, organizations, or objects). These  manifest as a primary key or foreign key reference. Examples: ``customers who,'' 
``products that,'' ``employees in.''

\item \textbf{WHAT} specifies record properties or characteristics (e.g., values, states, or calculated fields) and manifest as within-record field values or derived attributes.  Examples: ``with revenue over \$1M,'' ``colored blue,'' ``rated 5 stars.''

\item \textbf{WHERE} identifies location or scope (e.g., physical locations, logical boundaries, or containment hierarchies). This manifests as geographic or categorical fields.  Examples: ``in California,'' ``within department,'' ``belonging to category.''

\item \textbf{WHEN} indicates temporal considerations (e.g., time points, periods, sequences, or durations), manifesting as timestamps, date ranges, or temporal relationships. Examples: ``last quarter,'' ``before 2020,'' ``during business hours.'' Temporal constraints may additionally take derived values anchored to another entity rather than a fixed reference point, creating a WHO-anchored WHEN that requires subquery resolution rather than a simple filter predicate.

\item \textbf{WHY} describes causal relationships or explanatory connections between entities, manifesting as foreign key chains, association tables, or cross-schema linkages that encode why a record has a particular state. Examples: ``patients readmitted due to surgical complications," ``orders canceled because of stockouts," ``employees flagged for performance reasons." Of the six dimensions, WHY most directly approaches the boundary between retrieval and reasoning: a WHY constraint can sometimes be grounded in a recorded field, but often requires causal inference that SQL semantics do not natively support.

\item \textbf{HOW} identifies methods, processes, mechanisms, or procedural pathways, manifesting as process indicator fields, route or channel attributes, or aggregate functions when reducible to measurement. Examples: ``admitted through the emergency department," ``purchased via subscription," ``readmission rate by treatment arm." HOW queries that reduce to HOW MANY are expressible in standard SQL. HOW queries that ask by what mechanism an outcome occurs are not, marking HOW, like WHY, as a frontier dimension whose full treatment requires capabilities beyond current query systems.
\end{itemize}

\begin{figure}[t]
    \centering
    \includegraphics[width=0.9\textwidth]{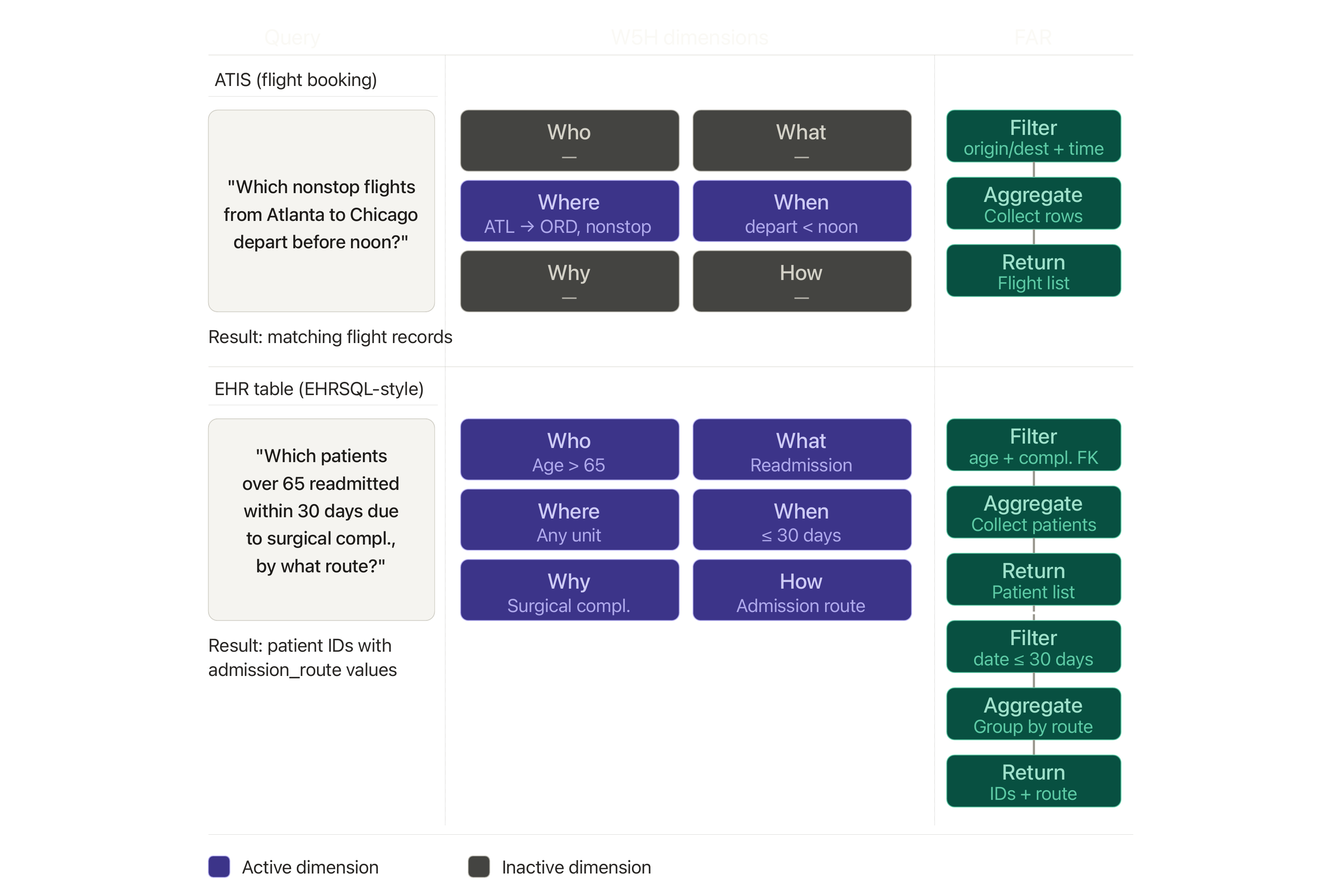}
    \caption{The QUEST translation pipeline for two representative queries from ATIS and EHRSQL. Each query maps from natural language (left) through W5H dimensional criteria (center) to FAR structural operations (right).}
    \label{fig:quest_overview}
\end{figure}

Figure \ref{fig:quest_overview} illustrates how these dimensions map across queries of varying complexity and domain.

\subsection{The FAR Structural Invariant}
\label{sec:far_theory}

While W5H characterizes the semantic content of queries, FAR characterizes their structure. Every well-formed database query reduces to three fundamental steps: \textbf{Filter} on criteria drawn from the W5H dimensions, \textbf{Aggregate} all records that meet those criteria, and \textbf{Return} the scoped result. This pattern holds regardless of domain, schema complexity, or query length. Simple single-table lookups and complex multi-table analytical queries alike reduce to the same Filter$\rightarrow$Aggregate$\rightarrow$Return structure. Complex queries decompose into sets of simpler queries, each conforming to the same pattern, by identifying the W5H criteria at each step and resolving them independently before combining the results. Figure \ref{fig:quest_overview} illustrates this decomposition across two queries of increasing complexity, including a multi-step case where the outer FAR structure depends on the resolved output of an inner FAR subquery.

\paragraph{Relationship to Relational Algebra.}
FAR is not independent of relational algebra, it is deliberately grounded in it. The Filter step corresponds to the selection operator ($\sigma$), the Aggregate step to projection and grouping operations ($\pi$, $\gamma$), and the Return step to the result relation passed to the caller. The contribution of FAR is not a mathematical discovery but a human-interpretable vocabulary for structural validation. It makes the preservation requirements of a query explicit in terms that are meaningful to developers and evaluators, not only to query optimizers. This grounding also defines what FAR nonconformance looks like in practice. A generated query that lacks any coherent filtering criterion (e.g., a query that returns an entire table without scoping) fails the Filter step regardless of whether it executes without error. A query that applies an aggregation unsupported by the schema (e.g., averaging a non-numeric field) fails the Aggregate step. A query whose SELECT clause returns columns unrelated to the user's intent fails the Return step. In each case, FAR nonconformance is detectable independently of execution outcomes, making it a complementary validation signal to execution accuracy rather than a restatement of it.

\paragraph{Constraint Operators and Logical Combinators.}
The FAR structure admits refinement without alteration. Constraint operators (e.g., \textit{cheapest, largest, earliest}) modify dimensional values within the Filter step, while logical combinators (AND, OR) join multiple dimensional constraints. Neither operators nor combinators change the underlying Filter$\rightarrow$Aggregate$\rightarrow$Return pattern. For example, a query filtering on the cheapest flight departing before noon is still a Filter$\rightarrow$Aggregate$\rightarrow$Return query; the constraint operator (cheapest) simply narrows the W5H specification.

\paragraph{The Filtered Enumeration Principle.}
Consider the query `which patients were readmitted last month?' The database receives a single instruction: find the records where readmission occurred within the specified time window (Filter), collect those records (Aggregate), and return them (Return). Every database query reduces to this same instruction. The Filtered Enumeration Principle states this formally. Every query is a command to enumerate or summarize (e.g., count, average, max, etc.) the records meeting a W5H dimensional specification. This reduction is universal. Complex queries decompose into sets of simpler queries that each conform to the same Filter$\rightarrow$Aggregate$\rightarrow$Return pattern, with explicitly identified W5H criteria at each step. Figure \ref{fig:quest_overview} illustrates this reduction across two queries of varying complexity.

\subsection{Case Studies}
\label{sec:case_studies}
The following three case studies illustrate the full QUEST translation pipeline, from natural language through W5H dimensional decomposition and FAR structure to SQL. The case studies demonstrate where existing approaches fall short and how QUEST diagnoses and resolves those failures. All examples use a simplified ATIS schema with a single table: 
\begin{Verbatim}[fontsize=\footnotesize]
FLIGHTS(flight_id, airline, origin_city, destination_city, departure_time, arrival_time, fare)
\end{Verbatim}

\subsubsection{Simple Query}
Consider the query: \textit{``Find all morning flights from Boston to New York.''}

\paragraph{W5H.} The query specifies WHERE twice: origin (Boston) and destination (New York),  and WHEN once (departure time before noon). No WHO, WHAT, WHY, or HOW dimensions are engaged. 

\paragraph{FAR.} Based on W5H, the FAR structure follows directly. Filter occurs on the WHERE and WHEN (origin, destination, and departure time). Aggregate is performed on the matching records and it returns the list of flights.

\paragraph{SQL.} The SQL query associated with the natural language query is:
\begin{Verbatim}[fontsize=\footnotesize]
SELECT *
FROM flights
WHERE origin_city = Boston
AND destination_city = New York
AND departure_time < noon
\end{Verbatim}

This case illustrates the baseline translation pipeline operating cleanly. W5H dimensions map directly to filter predicates, FAR structure maps directly to query operations, and the resulting SQL is unambiguous. Existing approaches handle this case correctly, which is precisely why simple queries are poor diagnostics for system capability.

\subsubsection{Semantic Ambiguity}
Existing approaches for text-to-SQL generally treat all temporal constraints as equivalent. As a result, they will produce correct SQL when temporal references are simple but might fail when they involve inter-entity relationships. Consider the query: \textit{``Find flights from Boston arriving before the last departure of flight AA100.''}

\paragraph{W5H.} This query contains two constraints that parsers may treat as the same type as both look like temporal filters. However, ``From Boston'' is a WHERE constraint on origin. ``Arriving before the last departure of flight AA100'' is a WHEN constraint, but one whose value is derived from another entity (flight AA100) making it a WHO-anchored WHEN. As a result, the two constraints require different SQL operations: a simple equality filter for WHERE, and a nested subquery for the WHO-anchored WHEN.

A text-to-SQL parser conflating the two constraints may produce:
\begin{Verbatim}[fontsize=\footnotesize]
SELECT *
FROM flights
WHERE origin_city = Boston
AND arrival_time < departure_time(AA100)
\end{Verbatim}

While this is syntactically plausible, it is semantically incomplete as it does not resolve the temporal anchor correctly. QUEST's W5H decomposition makes the dimensional distinction explicit, where the nested subquery for the WHO-anchored WHEN is:
\texttt{SELECT MAX(departure\_time) FROM flights WHERE flight\_id = AA100}.

\paragraph{FAR.} The FAR structure that follows the W5H is then: Filter on origin and the resolved temporal anchor; Aggregate matching records; Return the flight list. Exact match evaluation catches that the first translation is wrong, but cannot say why. W5H identifies the failure as a WHO-anchored WHEN constraint that requires subquery resolution, giving the developer a diagnostic and a fix.

\paragraph{SQL.} The correct SQL query associated with the natural language query is:
\begin{Verbatim}[fontsize=\footnotesize]
SELECT *
FROM flights
WHERE origin_city = Boston
AND arrival_time < (SELECT MAX(departure_time) FROM flights WHERE flight_id = AA100)
\end{Verbatim}

\subsubsection{Dimensional Conflation and Constraint Operator Misclassification}
Real queries often exhibit multiple simultaneous failure modes that existing approaches cannot distinguish. Consider the query: \textit{"Find the cheapest morning flight from Boston to New York that arrives at least 45 minutes before the next departing flight to London."}

\paragraph{W5H.}
A surface parser sees four constraints and treats them as equivalent 
filter predicates, producing the following naive translation:
\begin{Verbatim}[fontsize=\footnotesize]
SELECT *
FROM flights AS f1
WHERE f1.origin_city = Boston
AND f1.destination_city = New York
AND f1.departure_time < noon
AND f1.arrival_time <= (SELECT MIN(f2.departure_time) - 45 minutes
                        FROM flights f2
                        WHERE f2.destination_city = London)
AND f1.fare = cheapest
\end{Verbatim}
This fails in two independent ways. First, the correlated temporal constraint loses its correlation as the subquery returns the earliest London departure of the day rather than the next departure after this flight's arrival. Second, ``cheapest'' is misclassified as a filter predicate rather than a ranking operation on fare. Both failures produce syntactically valid SQL that executes without error but returns semantically wrong results. Exact match evaluation detects that the translation is wrong but cannot identify either failure mode, whether they are related, or how to correct either one.

W5H decomposition diagnoses both failures explicitly by revealing that the four constraints are dimensionally and operationally distinct:
\begin{itemize}
\item WHERE: origin\_city = Boston, destination\_city = New York
\item WHEN: departure\_time < noon (morning)
\item WHEN: arrival\_time $\leq$ next London departure - 45 minutes 
(a calculated temporal offset requiring subquery resolution)
\item Constraint operator: cheapest - a ranking operation on fare, 
not a filter predicate
\end{itemize}
The calculated WHEN constraint requires a correlated nested subquery. The constraint operator requires \texttt{ORDER BY} and \texttt{LIMIT} rather than a filter predicate. Each is an independent failure mode requiring an independent fix.

\paragraph{FAR.} The FAR structure then follows correctly: Filter on origin, destination, departure time, and resolved temporal offset; Aggregate and rank by fare; Return the top result.

\paragraph{SQL.} The correct SQL query associated with the natural language query is:
\begin{Verbatim}[fontsize=\footnotesize]
SELECT *
FROM flights AS f1
WHERE f1.origin_city = Boston
AND f1.destination_city = New York
AND f1.departure_time < noon
AND f1.arrival_time <= (SELECT MIN(f2.departure_time) - 45 minutes
                        FROM flights f2
                        WHERE f2.destination_city = London
                        AND f2.departure_time > f1.arrival_time)
ORDER BY f1.fare ASC
LIMIT 1
\end{Verbatim}

\section{Empirical Evaluation}
\label{sec:empirical}

To validate QUEST empirically, we apply it to five text-to-SQL datasets spanning multiple domains, query types, and levels of complexity. Our analysis pursues two questions: first, whether the FAR structural invariant (\textbf{Filter$\rightarrow$Aggregate$\rightarrow$Return}) holds universally as a structural property of well-formed queries; and second, whether the W5H dimensional profile varies systematically across domains in ways that have practical implications for system design and evaluation.

\subsection{Datasets}
\label{sec:datasets}

We analyze five publicly available text-to-SQL datasets representing
a range of domains, schema complexities, and query types:

\begin{itemize}
    \item \textbf{ATIS}~\citep{dahl1994atis} (Air Travel Information System; $n = 5{,}871$), a single-domain dataset of flight booking queries, representing a constrained vocabulary and schema with high query repetition.

    \item \textbf{WikiSQL}~\citep{zhong2017seq2sql} ($n = 80{,}654$), a large-scale crowdsourced dataset of queries over Wikipedia tables, representing broad topical coverage with relatively simple single-table schemas.

    \item \textbf{EHRSQL}~\citep{lee2022ehrsql} ($n = 24{,}405$), a healthcare dataset of natural language queries over electronic health record schemas derived from MIMIC-III and MIMIC-IV, representing the clinical deployment context most directly relevant to the motivating application.

    \item \textbf{Spider}~\citep{yu-etal-2018-spider} ($n = 8{,}034$), a cross-domain dataset with complex multi-table schemas spanning 200 databases across 138 domains, representing the  current standard for general-domain text-to-SQL evaluation.

    \item \textbf{BIRD}~\citep{li2023bird} (enterprise; $n = 1{,}500$), a dataset of queries over real-world enterprise database schemas with noisy data values and external knowledge requirements, representing complex analytical workloads at the upper end of query complexity.
\end{itemize}

Together these five datasets span the range from constrained single-domain queries (ATIS) through general-purpose cross-domain evaluation (Spider) to specialized healthcare (EHRSQL) and enterprise analytical (BIRD) workloads, providing a representative cross-section of the deployment contexts in which text-to-SQL systems operate ($n = 120{,}464$ total queries).

\subsection{FAR Conformance}
\label{sec:far_conformance}

We apply FAR structural analysis to all queries in each dataset, classifying each according to whether it exhibits filtering criteria (F), aggregation operations (A), and a return specification (R). A query is FAR-conformant if it reduces to the pattern of filtering on specified criteria, aggregating the resulting records, and returning the scoped result, the universal structural pattern posited by QUEST's Filtered Enumeration Principle. FAR conformance and W5H dimensional classification were evaluated using a large language model (Gemini, Google, December 2025) with a structured prompt specifying the framework definitions.

Every query in every dataset achieves 100\% FAR conformance across all 120,464 queries examined, single-table WikiSQL lookups, multi-table BIRD analytical workloads, and clinical EHRSQL queries alike, without exception. Because FAR is grounded in relational algebra, this universality is structurally expected: any well-formed SQL query will exhibit selection, grouping, and projection operations in some form. The contribution of this finding is therefore empirical confirmation that FAR provides a stable, domain-independent vocabulary for structural validation that applies uniformly across corpora without modification, and that the Filtered Enumeration Principle holds as a genuine universal property of database information retrieval, not an artifact of any particular domain or schema design.

This has two independent practical implications. First, FAR conformance is a reliable structural criterion for validating generated SQL independently of execution outcomes. A generated query that fails FAR conformance, whether by lacking a coherent filtering criterion, producing an unsupported aggregation, or returning a malformed result specification, is structurally incorrect regardless of whether it happens to return a non-empty result on a given test database. Second, because the Filtered Enumeration Principle is universal, the correctness of any translated query can be assessed against its canonical form, a command to enumerate or summarize the records meeting a W5H dimensional specification, independently of surface SQL syntax. This makes FAR a complementary validation signal to execution accuracy, not a substitute for it.

Within the FAR pattern, the three components play asymmetric roles. Return functions as a structural constant rather than a discriminating dimension: every well-formed query returns something, so conformance on this dimension is universal by construction. The analytical weight of the FAR pattern lies in Filter and Aggregate, which vary meaningfully across query types and domains. This asymmetry motivates richer taxonomies of query operations that decompose the structural space more finely.

\subsection{W5H Dimensional Analysis}
\label{sec:w5h}

\begin{figure}[t]
    \centering
    \includegraphics[width=\columnwidth]{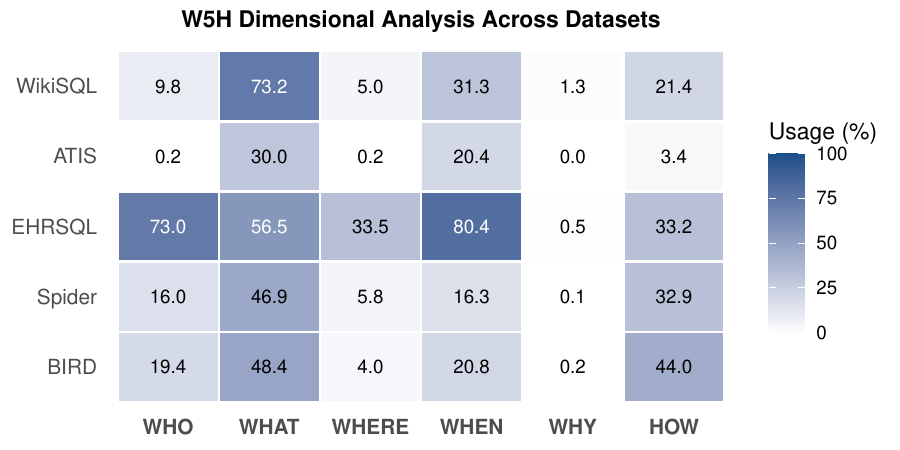}
\caption{Heatmap showing W5H dimensional profiles across five 
    text-to-SQL datasets. Rows represent the datasets; columns 
    represent the six dimensions (WHO, WHAT, WHERE, WHEN, WHY, 
    HOW). Color intensity reflects the proportion of queries 
    engaging each dimension, using a plasma palette. Note that 
    HOW values for BIRD (44.0\%) and Spider (32.9\%) predominantly 
    reflect quantitative HOW MANY queries (e.g., COUNT operations) 
    rather than mechanistic or procedural reasoning.}
    \label{fig:w5h_heatmap}
\end{figure}

Where FAR conformance is universal, the W5H dimensional profile of queries varies substantially and systematically across domains, the complementary finding that gives the framework its practical teeth. We classify each natural language query in each dataset according to which of the six dimensions (WHO, WHAT, WHERE, WHEN, WHY, HOW) its filtering criteria address, and compute the proportion of queries in each dataset that engage each dimension. Figure~\ref{fig:w5h_heatmap} displays the full dimensional profile across all five datasets as a heatmap.

The results reveal a striking pattern of domain differentiation. Four findings stand out.

\paragraph{Healthcare queries are dimensionally distinct from all other domains.} EHRSQL exhibits a distinctive dimensional profile that separates it from every other dataset examined. The WHEN dimension, identifying temporal constraints, date arithmetic, and time-bounded retrieval, appears in 80.4\% of EHRSQL queries, compared to 31.3\% in WikiSQL, 20.4\% in ATIS, 16.3\% in Spider, and 20.8\% in BIRD. The WHO dimension, identifying patients, providers, or other person-scoped entities, appears in 73.0\% of EHRSQL queries, compared to 9.8\% in WikiSQL, 0.2\% in ATIS, and 16.0\% in Spider. This concentration of temporal and person-centric dimensions reflects the fundamental structure of clinical reasoning: nearly every clinical query concerns a specific patient or population, bounded in time. No other domain in our analysis approaches this profile.

\paragraph{Aggregation density tracks query complexity.} Aggregation operations (e.g., COUNT, SUM, AVG, and related functions) appear in 38.3\% of EHRSQL queries, substantially above WikiSQL (23.4\%) and ATIS (4.1\%), and comparable to enterprise (BIRD: 43.4\%) workloads. This pattern confirms that healthcare queries share the computational complexity profile of enterprise analytical workloads, not the simpler lookup profile of general-domain benchmarks --- a finding with direct implications for benchmark selection in clinical text-to-SQL evaluation.

\paragraph{Causal reasoning is absent everywhere.} The WHY dimension, for causal and explanatory reasoning, is near-zero across all five datasets, appearing in fewer than 1\% of queries in every corpus examined. This is not a domain-specific gap but a universal characteristic of current text-to-SQL benchmarks: causal reasoning is not represented in any existing evaluation corpus at meaningful scale. This leaves system capability on this dimension entirely uncharacterized.

\paragraph{Mechanistic reasoning is equally absent.} The HOW dimension, for process, mechanism, and procedural queries, requires careful interpretation. The relatively moderate rates in BIRD (44.0\%) and Spider (32.9\%) predominantly reflect HOW MANY queries reducible to COUNT and other aggregate functions, a tractable subset that standard SQL handles directly. These figures therefore overstate system capability on genuine mechanistic reasoning: HOW queries that ask by what process or pathway an outcome occurs are essentially absent from all benchmarks examined, meaning systems that perform well on current benchmarks have simply never been evaluated on this dimension.

\subsection{Cross-Domain Comparison and Implications}
\label{sec:cross_domain}

The FAR and W5H results are complementary by design, and their juxtaposition is the central finding of this empirical analysis. FAR conformance is structural and universal: it holds without exception across all 120,464 queries, all five datasets, and all levels of schema complexity, establishing a common foundation for query analysis that does not depend on domain. W5H dimensional profiles are semantic and domain-specific: they vary substantially and systematically in ways that FAR conformance alone cannot detect and execution accuracy alone cannot characterize. A benchmark can achieve high FAR conformance and still be dimensionally mismatched to its target deployment domain, a gap that only becomes visible when the dimensional profile of the evaluation corpus is compared to that of the deployment domain.

\section{Discussion}
\label{sec:implication}
QUEST is not solely an analytical tool for characterizing existing queries. Its  two pillars have direct and complementary design implications: FAR, as a structural invariant, provides a validation criterion applicable at every stage of the text-to-SQL pipeline; W5H, as a semantic profiler, provides the dimensional vocabulary needed to align training data, benchmarks, and interfaces with target deployment domains. We describe five areas where these principles translate into actionable guidance.

\subsection{System Development and Evaluation}
\label{sec:txt2sql}

For text-to-SQL system developers and benchmark designers, QUEST provides two independent levers for improving reliability. The first is W5H dimensional profiling, which audits the semantic coverage of a training corpus or benchmark against the target deployment domain. A system trained primarily on WikiSQL or Spider and deployed in a clinical setting is not merely undertrained on edge cases. It is systematically undertrained on the dominant semantic operations its production queries will require, since the temporal and person-centric dimensions that define clinical reasoning appear at rates two to five times higher in EHRSQL than in any general-domain benchmark. Benchmark developers can apply the same profiling procedure in reverse: auditing dimensional coverage, identifying gaps, and generating targeted queries that probe underrepresented dimensions, producing evaluation sets that are more informative per query than random samples from existing distributions.

The second lever is FAR conformance checking, which provides an inexpensive structural screen for generated SQL independent of execution outcomes. A query that lacks a coherent filtering criterion, applies an unsupported aggregation, or returns a malformed result specification is structurally incorrect regardless of whether it happens to return a non-empty result on a finite test database. Because every well-formed query reduces to a command to enumerate or summarize the records meeting a W5H dimensional specification, the correctness of a translated query can also be assessed against this canonical form, a reference point that is more semantically grounded than execution accuracy alone and more tractable than unconstrained human judgment.

A third open question concerns the stability of system outputs under repeated invocations. Whether output reproducibility varies with W5H dimensional complexity, specifically whether queries engaging higher-order dimensions such as WHY and HOW produce less stable outputs across runs, remains uncharacterized. Recent work on reproducibility in biomedical text classification suggests that LLM outputs can vary meaningfully across repeated invocations even under identical inputs \citep{windisch2026onerun}, motivating analogous investigation in the text-to-SQL setting. QUEST's dimensional vocabulary provides a natural framework for stratifying such analysis by query complexity.
\subsection{Human-Facing Tools and Interfaces}
\label{sec:interface}

QUEST's two pillars translate directly into design principles for both end-user query interfaces and developer tooling. For non-technical users, W5H dimensions provide the natural organizing structure for query guidance, auto-completion, and error messaging. A user who specifies WHO and WHAT but omits WHEN can be prompted to consider whether temporal constraints are relevant, a prompt that is semantically meaningful to the user and computationally grounded in the dimensional structure of the query. FAR provides a complementary structural scaffold: because every well-formed query must filter, aggregate, and return, an interface can detect and surface structural incompleteness before query submission, addressing not `what are you asking about?' but `is your query well-formed?' For healthcare interfaces specifically, the concentration of WHEN and WHO in clinical queries suggests that temporal constraint specification and patient-scoping should be treated as first-class design elements, with date range selectors, temporal relation operators, and patient identifier fields surfaced prominently rather than buried in generic natural language input.

For developers, the same principles translate into concrete tooling capabilities. On the W5H side, dimensional tagging identifies which dimensions a query addresses and flags semantically underspecified queries, while dimensional coverage reporting characterizes the semantic profile of a test suite relative to a target deployment distribution. On the FAR side, conformance checking verifies the structural integrity of generated SQL before execution, and query decomposition assistance breaks complex queries into constituent simple queries each reducible to the Filter$\rightarrow$Aggregate$\rightarrow$Return pattern. Together these capabilities give developers visibility into the semantic structure of their query corpora that current text-to-SQL development practice does not provide.

\section{Related Work}
\label{sec:related}

\paragraph{Text-to-SQL Systems and Benchmarks.} The text-to-SQL problem has been studied extensively since the introduction of early datasets such as 
ATIS~\citep{dahl1994atis} and GeoQuery~\citep{zelle1996geoquery}. The field advanced substantially with WikiSQL ~\citep{zhong2017seq2sql}, which provided a large-scale crowdsourced benchmark, and Spider ~\citep{yu-etal-2018-spider}, which introduced cross-domain schema complexity and multi-table reasoning. More recent benchmarks, like EHRSQL~\citep{lee2022ehrsql} for electronic health records and Spider 2.0~\citep{lei2025spider20evaluatinglanguage} for enterprise workflows have pushed toward greater ecological validity. Despite this benchmark progression, evaluation methodology has remained anchored to two metrics: execution accuracy, which tests whether a predicted query returns the same result set as the gold standard on a test database, and exact set match, which tests structural identity. Both metrics measure average performance on known distributions without characterizing the structural or semantic operations 
a system must handle correctly. Pourreza and Rafiei~\citep{pourreza2023evaluating} conduct a systematic re-evaluation of top-performing models on prominent benchmarks and find that exact matching fails systematically due to query underspecification and reference query ambiguity --- a finding that directly motivates both the FAR structural validation and the principled query decomposition approach embedded in QUEST. Notably, Spider 2.0 documents a catastrophic performance collapse: models achieving 86\% execution accuracy on Spider drop to 6\% on Spider 2.0's enterprise workflows~\citep{lei2025spider20evaluatinglanguage}, revealing that semantic operations absent from standard benchmark distributions are precisely those required by real-world deployment --- the gap QUEST is designed to characterize and address.

\paragraph{Natural Language Interfaces to Databases.} The broader problem of natural language interfaces to databases (NLIDB) has a long history predating the deep learning era~\citep{androutsopoulos1995natural}. Early systems relied on hand-crafted grammars and semantic parsers; contemporary systems rely on fine-tuned language models and few-shot prompting. What has remained consistent across generations is the absence of a principled account of what semantic content a query expresses and what structural operations must be correctly handled for a translation to be faithful. QUEST provides both accounts: the FAR structural invariant characterizes the operations that must be preserved, while the W5H dimensional framework characterizes the semantic content that must be translated --- grounded in dimensions that are cognitively natural~\citep{lasswell1948structure} and formally tractable in relational algebra.

\paragraph{Query Understanding and Semantic Parsing.} Semantic parsing research has produced sophisticated approaches to mapping natural language to formal 
representations, including lambda calculus~\citep{zettlemoyer2005learning}, 
dependency-based compositional semantics~\citep{liang2013lambdadependencybasedcompositionalsemantics}, and abstract meaning representations~\citep{banarescu2013amr}. These frameworks are designed for broad-coverage semantic representation rather than for characterizing the preservation requirements of database query systems specifically. QUEST occupies a complementary niche: rather than a general semantic representation language, it is a principled decomposition of the structural and semantic axes along which database queries vary, designed to support systematic evaluation of whether text-to-SQL systems preserve query intent across both dimensions.

\paragraph{Healthcare NLP and Clinical Query Systems.} Clinical text-to-SQL systems face distinctive evaluation challenges rooted in the semantic structure of clinical reasoning. The i2b2/n2c2 shared task series~\citep{uzuner20112010,henry20202018} has produced annotated datasets and evaluation protocols for clinical concept extraction, assertion classification, and temporal reasoning spanning nearly two decades of community effort. These protocols are rigorous within their task definitions but are not designed to characterize the dimensional profile of semantic variation that clinical queries exhibit in deployment. Our empirical analysis establishes that healthcare queries have a distinctive semantic fingerprint, with temporal (WHEN) and person-scoped (WHO) dimensions appearing in 80.4\% and 73.0\% of EHRSQL queries respectively, far exceeding general-domain benchmarks --- a gap that QUEST's dimensional analysis makes visible and measurable, and that FAR conformance checking alone would leave entirely undetected.

\section{Conclusion}
\label{sec:conclusion}

We presented QUEST (Query Understanding Evaluation through Semantic Translation), a principled framework for database information retrieval built on two co-equal pillars: the FAR structural invariant and the W5H dimensional framework, unified by the Filtered Enumeration Principle. Just as structured programming brought systematic discipline to software development, QUEST brings the same reductive insight to database queries, demonstrating that all queries reduce to a universal Filter $\rightarrow$\ Aggregate $\rightarrow$ Return pattern applied to filtering criteria drawn from six semantic dimensions.

The empirical validation establishes two findings of independent significance. FAR conformance is genuinely universal: every well-formed query in every corpus we examined reduces to the Filter $\rightarrow$ Aggregate $\rightarrow$ Return pattern without exception across 120,464 queries and five datasets spanning multiple domains and schema types. W5H dimensional profiles are domain-specific in ways that matter: healthcare queries exhibit a distinctive semantic fingerprint, concentrated in temporal and person-centric dimensions at rates two to five times higher than general-domain benchmarks, a gap that standard evaluation methodology neither detects nor addresses. A text-to-SQL system evaluated on general-domain benchmarks and deployed in a clinical setting is being asked to perform semantic operations its evaluation never probed.

Beyond these contributions, QUEST points toward a frontier that is among the most consequential open problems in the field. Causal and mechanistic reasoning, the WHY and HOW dimensions, are absent from all existing text-to-SQL benchmarks at meaningful scale. A system that correctly handles WHO, WHAT, WHERE, and WHEN is a sophisticated retrieval system; a system that correctly handles WHY and HOW is beginning to reason. Crossing that boundary requires benchmark development that explicitly labels causal and mechanistic queries, schema extensions that represent causal structure alongside tabular data, and evaluation oracles that go beyond execution accuracy. 

QUEST makes two things possible that were not possible before: characterizing the structural properties of any query corpus through FAR conformance analysis, and characterizing its semantic properties through W5H dimensional profiling. Together they provide the vocabulary and analytical procedure needed to close the gap between benchmark performance and deployment reliability, and to make progress toward the WHY and HOW frontier visible and measurable rather than merely aspirational.

\bibliographystyle{plainnat}
\bibliography{grammar}

\end{document}